\documentclass[aps,prr,superscriptaddress,longbibliography,twocolumn,showpacs,noeprint]{revtex4-2}
\usepackage{lipsum}
\usepackage{amssymb}
\usepackage{braket}
\usepackage{graphicx}
\usepackage{mathrsfs}
\usepackage{bm}
\usepackage{tabularx}
\usepackage{amsmath}
\usepackage[colorlinks=true,linkcolor=blue,citecolor=blue,urlcolor=blue]{hyperref}
\usepackage{tikz} %textcolor
\DeclareMathOperator{\Tr}{Tr}

\newcommand{\RN}[1]{%
	\textup{\uppercase\expandafter{\romannumeral#1}}%
}
\newcommand{\Rn}[1]{%
	\textup{\lowercase\expandafter{\romannumeral#1}}%
}

\begin{document}

	\title{Universal Properties of the Spectral Form Factor in Open Quantum Systems}
	\author{Yi-Neng Zhou}
	\affiliation{Institute for Advanced Study, Tsinghua University, Beijing,100084, China}

	\author{Tian-Gang Zhou}
	\affiliation{Institute for Advanced Study, Tsinghua University, Beijing,100084, China}

	\author{Pengfei Zhang}
 	\thanks{PengfeiZhang.physics@gmail.com}
	\affiliation{Department of Physics, Fudan University, Shanghai, 200438, China}
	\date{\today}

\begin{abstract}
The spectral form factor (SFF) can probe the eigenvalue statistic at different energy scales as its time variable varies. In closed quantum chaotic systems, the SFF exhibits a universal dip-ramp-plateau behavior, which reflects the spectrum rigidity of the Hamiltonian. In this work, we explore the universal properties of SFF in open quantum systems. We find that in open systems the SFF first decays exponentially, followed by a linear increase at some intermediate time scale, and finally decreases to a saturated plateau value. We derive universal relations between (1) the early-time decay exponent and Lindblad operators; (2) the long-time plateau value and the number of steady states. We also explain the effective field theory perspective of universal behaviors. We verify our theoretical predictions by numerically simulating the Sachdev-Ye-Kitaev (SYK) model, random matrix theory (RMT), and the Bose-Hubbard model. 
	
\end{abstract}
	
\maketitle
	
\emph{Introduction.} The spectral form factor (SFF) has attracted much attention in recent years for its direct relation to the eigenvalue statistics at different energy scales and its utility as a robust diagnosis of quantum chaos\cite{brezinUniversalityCorrelationsEigenvalues1993,brezinCorrelationsNearbyLevels1996,liuSpectralFormFactors2018,mullerPeriodicorbitTheoryUniversality2005,cotlerBlackHolesRandom2017,kosManyBodyQuantumChaos2018,bertiniExactSpectralForm2018a,chanSpectralStatisticsSpatially2018,kudler-flamConformalFieldTheory2020,royRandomMatrixSpectral2020,winerHydrodynamicTheoryConnected2022a,roySpectralFormFactor2022,barneySpectralStatisticsMinimal2023}. The structure of SFF is a direct indicator of the energy spectrum correlation in quantum systems. As its time variable increases, it reveals the eigenvalue statistics at a smaller energy scale. The SFF among different models reveals the symmetry that those models preserve. It exhibits several universal properties including the initial decay, the increase at intermediate time scales which shows a linear ramp in models with spectrum rigidity, and finally the saturation to a plateau value. This "dip-ramp-plateau" structure is ubiquitous in quantum chaos systems\cite{saadSemiclassicalRampSYK2019a,gharibyanOnsetRandomMatrix2018,winerExponentialRampQuadratic2020}. 
 
However, the interaction and exchange between the system and environment are inevitable, and therefore it is natural to focus on the corresponding problem in open systems. Generalizing familiar concepts in closed systems to open systems has helped people discover much more interesting and novel physics. Recently, the development of entropy dynamics\cite{zhouEnyiEntropyDynamics2021a}, entanglement phase transition\cite{mazzucchiQuantumMeasurementinducedDynamics2016,liQuantumZenoEffect2018,skinnerMeasurementInducedPhaseTransitions2019,liMeasurementdrivenEntanglementTransition2019,szyniszewskiEntanglementTransitionVariablestrength2019,chanUnitaryprojectiveEntanglementDynamics2019,vasseurEntanglementTransitionsHolographic2019,zhouEmergentStatisticalMechanics2019,gullansScalableProbesMeasurementInduced2020,jianMeasurementinducedCriticalityRandom2020,fujiMeasurementinducedQuantumCriticality2020,zabaloCriticalPropertiesMeasurementinduced2020,gullansDynamicalPurificationPhase2020,choiQuantumErrorCorrection2020,baoTheoryPhaseTransition2020,nahumMeasurementEntanglementPhase2021,PhysRevB.103.174309,sangMeasurementprotectedQuantumPhases2021,albertonEntanglementTransitionMonitored2021,lavasaniMeasurementinducedTopologicalEntanglement2021,PhysRevB.103.224210,LeGal:2022rwf,jianMeasurementInducedPhaseTransition2021a,zhangUniversalEntanglementTransitions2022,liuNonunitaryDynamicsSachdevYeKitaev2021,zhang2021emergent,zhang2022quantum,PhysRevB.106.224305}, operator complexity\cite{liuKrylovComplexityOpen2022,bhattacharyaOperatorGrowthKrylov2022,PathakOperatorGrowthOpen2023,bhattacharyaKrylovComplexityOpen2023} in open quantum many-body systems have aroused much interest in condensed matter physicists. Also, there are explorations on the definitions and physical interpretations of the SFF in open systems or non-Hermitian Hamiltonians\cite{liSpectralStatisticsNonHermitian2021c,kosChaosErgodicityExtended2021a,kawabataDynamicalQuantumPhase2022,xuThermofieldDynamicsQuantum2021,corneliusSpectralFilteringInduced2022a,matsoukas-roubeasNonHermitianHamiltonianDeformations2023}.
	
In this paper,  we study the SFF in open quantum systems driven by the Lindblad master equation. The definition we use excludes possible exponential growth over time in general non-Hermitian systems. For concreteness, we define the normalized SFF as the ratio between SFF with dissipation and that without dissipation. We find some universal properties of the normalized SFF according to its early-time and late-time dynamics. More specifically, we find this normalized SFF has an early-time exponential decay behavior related to the Lindblad operators and late-time plateau behavior related to the number of steady states. We demonstrate the universality of these properties in open systems by studying three different models with dissipation: the random matrix model, the SYK model, and the Bose-Hubbard model. In the random matrix model and the Bose-Hubbard model, the numerical results agree well with our conjecture. Furthermore, using the path-integral method, we give a candidate semi-classical explanation of the SFF in systems with dissipation, and this is a novel perspective for understanding the universal properties of the normalized SFF.
\begin{figure}[t] 
	\centering 
	\includegraphics[width=0.5\textwidth]{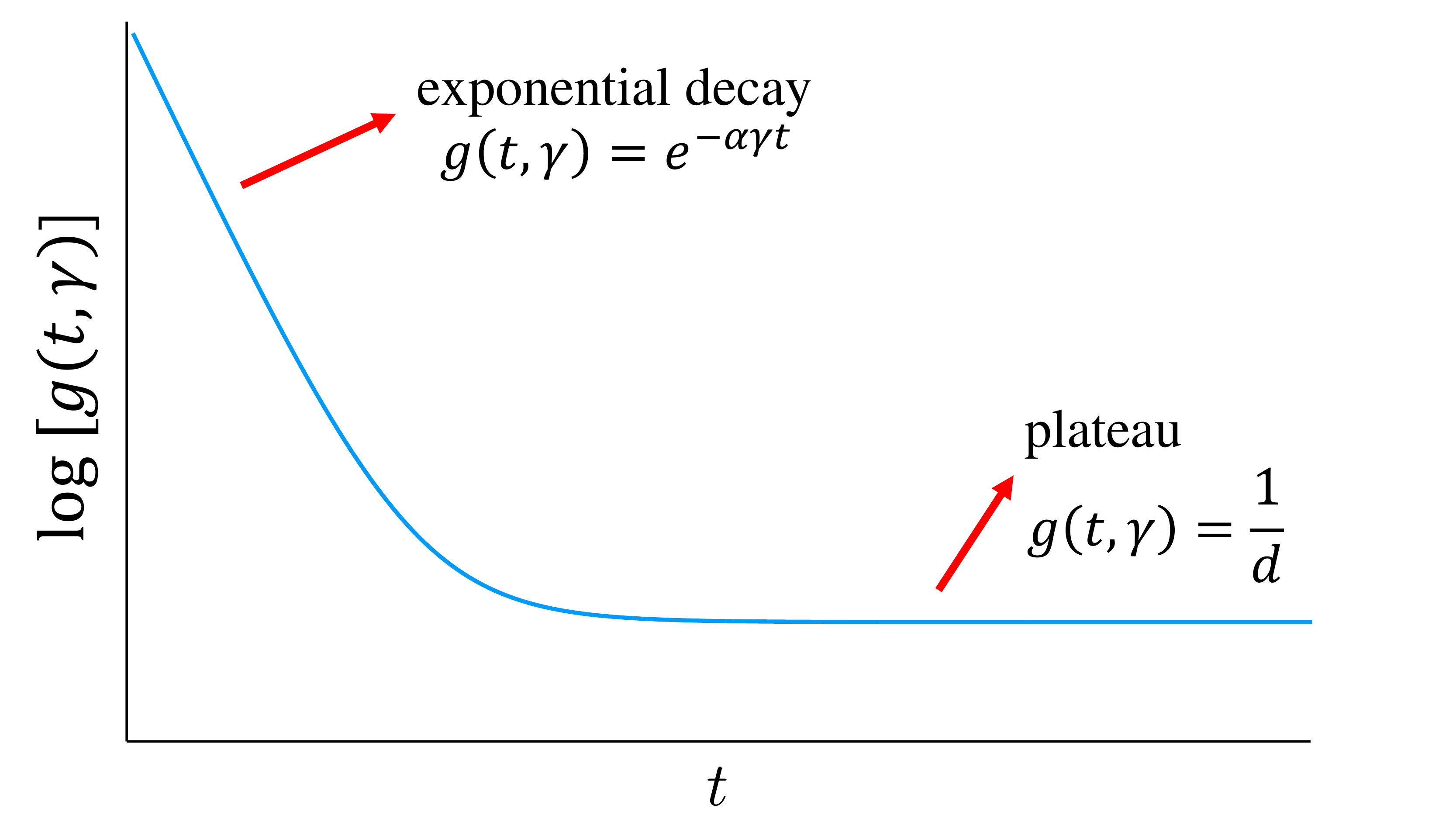} 
	\caption{The universal properties of the normalized SFF in open systems. Its has an early-time exponential decay and a long-time plateau behavior.}
	\label{cartoon}
\end{figure}

\emph{The definition of SFF in open systems.} In the closed system, the SFF can be defined as the size fluctuation of the analytic continuation of the thermal partition function of the quantum system
\begin{equation} 
	\begin{split}
		&F(t,\beta=0)=\frac{|\mathcal{Z}(it)|^2}{[\mathcal{Z}(0)]^2}=\frac{1}{[\mathcal{Z}(0)]^2}\sum_{m,n} e^{-i(E_m-E_n)t}.
		\label{close_SFF_definition}
	\end{split}
\end{equation}
with $\mathcal{Z}(it)=\operatorname{Tr}(e^{-itH})$.
From this expression, we see that SFF captures the energy level correlations of the full spectrum of the system, and the energy scale that it probes decreases as its time variable increases. At early time, SFF captures the energy level correlations at an energy scale much larger than the mean energy level spacing of the system, and it usually has a decay behavior, often called \textit{slope}. This slope region is non-universal in different models for it sees the details of the energy spectrum of the system. At the intermediate time scale, SFF measures the energy level correlation in the same order as the mean energy level spacing, and in some models that have level repulsion, we see a linear ramp of SFF as time increases. Therefore, SFF can be used to diagnose spectral rigidity. Over a long-time, the SFF often saturates to a constant plateau value determined by each single energy level.  Also, there are some studies about the non-universal properties of the form factor in chaotic systems\cite{agamSpectralStatisticsDisordered1995,bogomolnyGutzwillerTraceFormula1996a}. 

In the open system, we consider the time evolution of the system driven by the Lindblad Master equation 
	\begin{equation} 
		\frac{\partial{\rho}}{\partial t}=-i[H,\rho] +2\gamma\sum_m L_m \rho L^{\dagger}_m-\gamma\sum_m\lbrace L^{\dagger}_m L_m, \rho\rbrace.
	\end{equation}
Here, $\gamma$ is the dissipation strength, and $L_{\alpha}$ is the Lindblad jump operator. 
If we use the Choi-Jamiolkwski isomorphism\cite{TysonOperatorSchmidtDecompositionsFourier2003,VidalMixedStateDynamicsOneDimensional2004} to map the density matrix $\rho = \sum_{m,n} \rho_{mn}|m\rangle \langle n|$ to a wave function defined on a double space as $| \psi_{\rho}^D(t)\rangle=\sum_{mn}\rho_{mn}|m\rangle_L\otimes |n\rangle_R$, then after this mapping the wave function $\psi_{\rho}^D$ in the double system satisfies a Schrodinger-like equation $	i \partial_t{\psi_{\rho}^D(t)}=H^D\psi_{\rho}^D(t)$.
Here, $H^D=H_s-iH_d$ is defined on the double space with 
\begin{equation} 
\begin{split}
H_s &= H_L\otimes \mathcal{I}_R - \mathcal{I}_L\otimes H^T_R,\\
	H_d &=\gamma \sum_m [-2\hat{L}_{m,L}\otimes \hat{L}_{m,R}^*
 \\&+(\hat{L}^\dagger_m\hat{L}_m)_L\otimes \mathcal{I}_R +\mathcal{I}_L\otimes (\hat{L}^\dagger_m\hat{L}_m)_{R}^*].
 \end{split}
    \label{Hd}
\end{equation}
Operators with subscript $L$ and $R$ stand for operators acting on the left and the right systems respectively, and $T$
stands for the transpose, and $\mathcal{I}$ represents the identity operator. Here, $H_L$ and $H_R$ both takes the same form as the original Hamiltonian $H$, although they operators on different Hilbert space.
 
Similar to the SFF defined in the closed system in the Eq.~\eqref{close_SFF_definition}, we can define the  SFF in the open system as 
	\begin{equation} 	
	F_\gamma(t) =\frac{1}{[\mathcal{Z}(0)]^2} \operatorname{Tr}(e^{-iH^D t})=\frac{1}{[\mathcal{Z}(0)]^2}\sum_l   e^{(-i\alpha_l-\beta_l)t} .
	\label{definition_2_beta0}
	\end{equation}
Here, we use the subscript $\gamma $ to denote the SFF in open systems (that is the dissipation strength $\gamma$ is non-zero).  When we set the dissipation strength in the Lindblad evolution as zero, we find that this definition is the same as that in the closed system Eq.~\eqref{close_SFF_definition}.

Since the imaginary part of the Lindblad spectrum is always non-negative, this SFF defined in Eq.~\eqref{definition_2_beta0} will not grow exponentially. Thus, although the Lindblad spectrum is complex, the SFF defined in Eq.~\eqref{definition_2_beta0} will decay exponentially in time till it reaches the steady state value. In addition, there is an alternative approach to defining the SFF in open systems that has a close relation to the definition the Eq.~\eqref{definition_2_beta0}, and the details of this discussion are included in the supplementary material\cite{SM}.

\emph{The universal function of the normalized SFF.} Let us now consider the behavior of the normalized SFF in open systems defined as
\begin{equation} 
	g(t,\gamma)  \equiv\frac{	F_\gamma(t) }{	F(t,\beta=0) }.
	\label{normalized_SFF}
\end{equation}
The motivation here is to find some universal properties of this normalized SFF.
We summarize some universal properties of this normalized SFF including the early-time exponential decay behavior related to the Lindblad operators and the late-time plateau behavior related to the number of the steady state, and it is illustrated in Fig.~\ref{cartoon}. We summarize these universal properties below:

1.At the early time $\gamma t \ll 1$, the normalized SFF has an exponential decay behavior 
\begin{equation} 
	g(t,\gamma)  =e^{-\alpha \gamma t}.\ \ \ \text{with}\ \alpha = \sum_m 2\langle L_m^{\dagger}L_m\rangle_c
	\label{alpha_def}
\end{equation}
Here $\langle A\rangle\equiv \operatorname{Tr}\left[ A\right]/d$ and $\langle A B\rangle_c \equiv \langle A B\rangle-\langle A \rangle\langle B\rangle$. $d$ is the Hilbert space dimension of the Hamiltonian $H$.

2. The long-time behavior of this normalized SFF is a constant plateau whose value is given by 
\begin{equation} \label{longtime_SFF}
	g(t\to \infty,\gamma \neq 0)  = \frac{1}{d}.
\end{equation}

Below, we give some simple arguments for these universal properties. At small $\gamma t$, the SFF becomes
\begin{equation} 
	\begin{split}
		&F_\gamma(t)
		\simeq\frac{1}{[\mathcal{Z}_0(0)]^4}\operatorname{Tr}[  e^{-iH_st}]\operatorname{Tr}[e^{-H_dt}]\\
		&=F(t,\beta=0)\frac{1}{[\mathcal{Z}_0(0)]^2}\operatorname{Tr}[e^{-H_dt}].
	\end{split}
	\label{initial_alpha}
\end{equation}
In the early-time regime, it is known that the correlation between the left and right contour is much smaller than the correlation within the same contour \cite{saadSemiclassicalRampSYK2019a}, then we ignore the correlation contribution of the first term of the $H_d$ in Eq.~\eqref{Hd} when evaluating the last line of Eq.~\eqref{initial_alpha}. Using the fact that the second and the third terms of the $H_d$ commute with each other, we further obtain 
\begin{equation}
	F_\gamma(t)\simeq F(t,\beta=0)e^{- \sum_m 2\langle L_m^{\dagger}L_m\rangle_c\gamma t}.
\end{equation}
This leads to the expression of $\alpha$ in the Eq.~\eqref{alpha_def}, and its detailed derivation is in the supplementary material\cite{SM}. As time increases, the correlation between the left and right contour generally increases, thus the assumption above is not valid at the intermediate time. Therefore, the normalized SFF generally does not have this exponential decay behavior at the intermediate time scales $\gamma t \sim 1$.

The final plateau value of the normalized SFF can be understood by investigating Eq.~\eqref{definition_2_beta0}. Only the steady state with zero-imaginary eigenvalue will give a non-vanishing contribution to the long-time plateau value of SFF, and this gives the expression Eq.~\eqref{longtime_SFF}. In addition, if there are more than one steady state, then Eq.~\eqref{longtime_SFF} should be changed to 
$g(t\to \infty,\gamma \neq 0)  = \frac{\theta}{d}$. Here, $\theta$ is the total number of steady states.

Moreover, we can analyze the late-time regime using the effective field theory approach\cite{saadSemiclassicalRampSYK2019a,winerHydrodynamicTheoryConnected2022a}. The main idea is to approximate Green's functions on the path-integral contour of the SFF by their counterparts on a Keldysh contour with an auxiliary imaginary time separation between forward and backward evolutions. The SFF defined in the Eq.~\eqref{definition_2_beta0} can be written as the path-integral
\begin{equation}
    F_{\gamma}(t)=\int \mathcal{D}\Delta \mathcal{D} E_{\text{aux}} e^{-S_{\text{eff}}(\Delta,E_{\text{aux}})}.
\end{equation}
Without any dissipation, the linear ramp can be understood as an integration over the zero mode $\Delta$ and its conjugate variable $E_{\text{aux}}$. $\Delta$ describes the relative time shift between forward and backward evolution branches and $E_{\text{aux}}$ can be understood as the energy of the system. In closed systems, there is no coupling between two branches, and The effective action $S^0_\text{eff}(\Delta,E_{\text{aux}})$ does not depend on $\Delta$. Consequently, the integral over $\Delta$ from $0$ to $t$ leads to a linear slope. When the dissipation strength becomes small but finite, we find perturbatively:
\begin{equation}
    \delta S_\text{eff}=-2\gamma t \sum_i G_{\text{W},i}(\Delta,E_{\text{aux}}).
\end{equation}
Here $G_{\text{W},i}(t,E_{\text{aux}})$ is the Wightmann Green's function of operator $L_i$ with energy $E_{\text{aux}}$ \cite{saadSemiclassicalRampSYK2019a}
\begin{equation}
    G_{\text{W},i}(\Delta,E_{\text{aux}})=\langle e^{iH\Delta}e^{-\frac{\beta_{aux} }{2}H}L_i e^{-iH\Delta}e^{-\frac{\beta_{aux} }{2}H}L_i^\dagger \rangle_c,
\end{equation}
where $\beta_{\text{aux}}(E_{\text{aux}})$ is determined by the thermodynamic relation.
 This leads to a finite mass for $\Delta$, which increases linearly as time increases. In particular, as $t\rightarrow \infty$, the mode will be pinned at $\Delta =0$, which terminates the presence of the linear ramp.

\emph{Examples.} In the following, we use the SYK model, the random matrix model, and the Bose-Hubbard model as examples to illustrate these universal properties of the normalized SFF in the open system. 

We comment here that the SYK model and the random matrix model are both good examples to analytically calculate the SFF since they both involve random averages over different realizations that rattle the energy eigenvalues.  The random average smooths out the fluctuations that come from the oscillating terms in the SFF, thus making it a smooth function of time. In comparison, the SFF has extensive spikes in the Bose-Hubbard model that come from the zeros of the SFF, and we need to do the time slice average to get a smooth  SFF curve.

\emph{A. SYK Model.} 
We consider the SFF of the SYK model whose Hamiltonian is of the form
\begin{equation} 
	H = i^{\frac{q}{2}}\sum_{a_1<...<a_q}^N J_{a_1,...,a_q}\psi_{a_1}...\psi_{a_q}.
\end{equation}
Here, $J_{a_1,...,a_q}$ is a random variable that satisfies the Gaussian
distribution with mean zero and variance 
\begin{equation*} 
	\langle J_{a_1,...,a_q} J_{a_1^{'},...,a_q^{'}} \rangle=\delta_{a_1,a_1^{'}}...\delta_{a_q,a_q^{'}}\frac{J^2(q-1)!}{N^{q-1}},
\end{equation*}
and $\psi$ is the Majorana fermion operator.

\begin{figure}[t] 
	\centering 
	\includegraphics[width=0.5\textwidth]{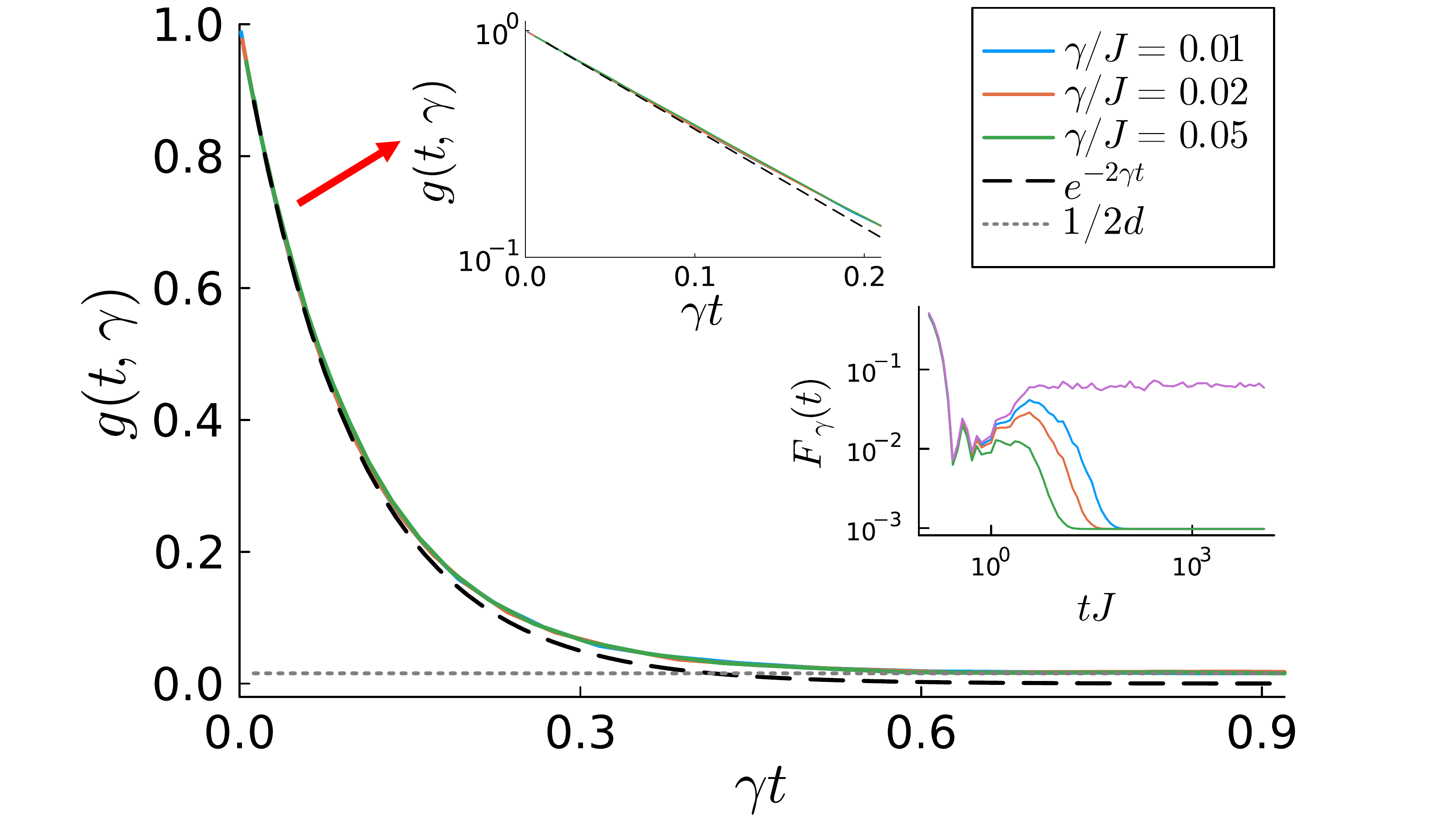} 
	\caption{The normalized SFF dynamics for SYK model as a function of $\gamma t$. $\gamma$ is the dissipation strength. The Lindblad jump operators are chosen as the single Majorana Fermion operators. The dashed line is a theoretical prediction of the initial slope based the Eq.~\eqref{alpha_def}. The gray dot line is the prediction of the final plateau value. The left inset shows the early-time behavior of the normalized SFF. The right inset is a log-log plot of the SFF at different dissipation strengths, and the purple line is SFF without dissipation for comparison. The total number of Majorana Fermion is $N=10$, and the random sample sizes is 200.}
	\label{SYK_all}
\end{figure} 
We numerically compute the SFF in Fig.~\ref{SYK_all}, and there are several noteworthy features of this figure. First, we find that curves with different dissipation $\gamma$ collapse well into a single line when they are plotted in terms of $\gamma t$. Second, the early-time exponential decay in the SYK model is visible in the Fig.~\ref{SYK_all}, and it agrees well with our analytical result $e^{-N\gamma t}$ at early time region $\gamma t<0.2$. Third, the long-time value of the SFF curve is a non-vanishing plateau whose value is $1/2^{N+1}$.

Furthermore, we can then write the SFF of the SYK model as a path-integral with the Lindblad operator chosen as the single Majorana fermion operator $L_i = \psi_i$. Also, the dissipation strength is chosen as the constant $\gamma$. We can then solve the early-time saddle-point solutions of the effective action, and to the first-order of dissipation strength $\gamma$, the effective action at the saddle point is $I[G,\Sigma]=I_0[G,\Sigma]+N\gamma T$. Thus, we obtain the normalized SFF as $g(t,\gamma) =  \exp[-N\gamma T]$. It has an exponential decay behavior at the early time. The details of the derivation of the SFF in the SYK model are included in the supplementary. A similar analysis of the SFF in the Brownian SYK is also included, in which the normalized SFF also has an early-time exponential decay behavior\cite{SM}. Moreover, since the spectrum of Majorana SYM model with N mod 8 is not 0 has a 2-fold degeneracy\cite{youSachdevYeKitaevModelThermalization2017,cotlerBlackHolesRandom2017}, thus the final plateau value of the normalized SFF is $\frac{1}{2d}$ instead of $\frac{1}{d}$ as shown in the Fig.~\ref{SYK_all}.

\emph{B. The Random Matrix Theory.}  Consider the SFF in Gaussian unitary ensemble (GUE). The SFF we defined in Eq.~\eqref{definition_2_beta0} can be written in RMT as 
\begin{equation} 
	F_\gamma(t)=\frac{1}{N^2}\langle \operatorname{Tr} [e^{iH^Dt}] \rangle
\end{equation}
with $H^D=H_s -i H_d$ being a random matrix defined on double space. Here, $H_s = H_L\otimes \mathcal{I}_R - \mathcal{I}_L\otimes H^T_R$, and $H_d =\gamma (-2\hat{L}_L\otimes \hat{L}_R^*+(\hat{L}^\dagger\hat{L})_L\otimes \mathcal{I}_R +\mathcal{I}_L\otimes (\hat{L}^\dagger\hat{L})_R^*$ with $H$ and $L$ both are $N \times N $ random Hermitian matrix. The bracket $\langle \cdots \rangle$ means an averaging with respect to the Gaussian distribution:
\begin{equation} 
	P(H)=\frac{1}{\mathcal{Z}}e^{-\frac{N}{2}\operatorname{Tr}(H^2)},\ \ \ P(L)=\frac{1}{\mathcal{Z}}e^{-\frac{N}{2}\operatorname{Tr}(L^2)}.
\end{equation}
Then we consider the SFF in open systems in RMT. The SFF of open systems defined in Eq.\ref{definition_2_beta0} can be written as 
\begin{equation} 
	F_\gamma(t) =\frac{\int dH dLe^{-\frac{N}{2}\Tr H^2}e^{-\frac{N}{2}\Tr L^2}\Tr e^{-itH^D}}{\int dH dL e^{-\frac{N}{2}\Tr H^2}e^{-\frac{N}{2}\Tr L^2}}.
\end{equation}
\begin{figure}[t] 
	\centering 
	\includegraphics[width=0.53\textwidth]{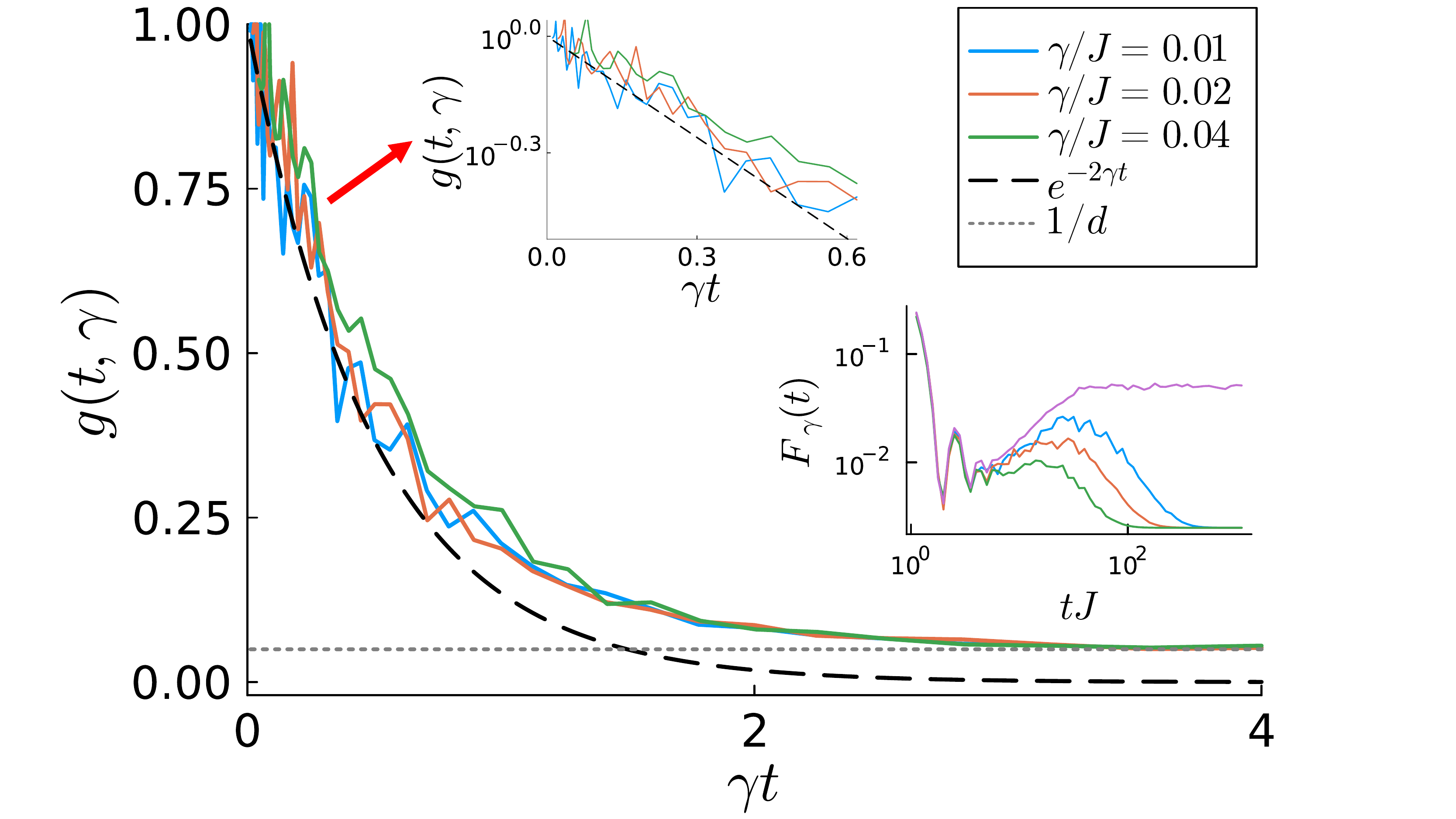} 
	\caption{The normalized SFF dynamics for GUE random matrices with dimension $N_{dim}=20$ as a function of $\gamma t$. $\gamma$ is the dissipation strength. The Lindblad jump operators are chosen as the random hermitian matrix of GUE. The dashed line is a theoretical prediction of the initial slope based the Eq.~\eqref{alpha_def}. The gray dot line is the prediction of the final plateau value. The left inset shows the early-time behavior of the normalized SFF. The right inset is a log-log plot of the SFF as a function of $tJ$, and the purple line is SFF without dissipation for comparison. Here, the random realization of $H$ and $L$ is independent, and we randomize them each for 100 realizations.}
	\label{RMT_compare}
\end{figure}

In Fig.~\ref{RMT_compare}, we present $F_\gamma(t)$ for the GUE ensemble of matrices with dimension $N=20$. We find that without dissipation the SFF first dips below its plateau value and then climb back up in a linear fashion (this region is also called the \textit{ramp}), joining onto the plateau as depicted in the right inset of the Fig.\ref{RMT_compare}. Also, when we add a small dissipation, we find a similar dip-ramp behavior of the SFF, whereas it then decays to a plateau value that is lower than the case without dissipation. Moreover, the height of the plateau is of order $1/N$ without dissipation which is the mean level spacing, and the height of the plateau is of order $1/N^2$ with non-zero dissipation. 
	
To understand this behavior of SFF with dissipation, we can directly calculate the normalized SFF, and the derivation details are included in the supplementary material\cite{SM}. We obtain the normalized SFF at early times 
\begin{equation} 
	g(t,\gamma)\simeq e^{-2\gamma t}, \gamma t \ll 1,
\end{equation}
and this is an exponential decay behavior which is also visible in the numerical results in Fig.~\ref{RMT_compare}, and it is in good agreement with $e^{-2\gamma t}$ at $\gamma t<0.5$.
On the other hand, in the long time limit $t \to \infty$, we find $g(t\to \infty,\gamma \neq 0)=\frac{1}{N}$, and $g(t\to \infty,\gamma =0)=1$. This explains the difference between the final plateau value in the case with and without dissipation as depicted in Fig.~\ref{RMT_compare}.

\emph{C. Bose-Hubbard Model.} We now consider the SFF in the 1D Bose-Hubbard model with dissipation. The Hamiltonian of the Bose Hubbard model is
\begin{equation}
	\hat{H}=-J\sum_{\langle i,j\rangle}\hat{b}_{i}^{\dagger}\hat{b}_{j}+\frac{U}{2}\sum_{i}\hat{n}_{i}(\hat{n}_{i}-1)
\end{equation}
Here, $J$ is the strength of the nearest neighbor hopping, and $U$ is the strength of the on-site interaction. In an open system, we set $\gamma$ as a time-independent dissipation strength. Also, we set the Lindblad jump operators as $\hat{L}_{m}=\hat{n}_{m}$. Here $m=1,2,..., N_s$, and $N_s$ is the total number of sites. 
\begin{figure}[t] 
	\centering 
	\includegraphics[width=0.52\textwidth]{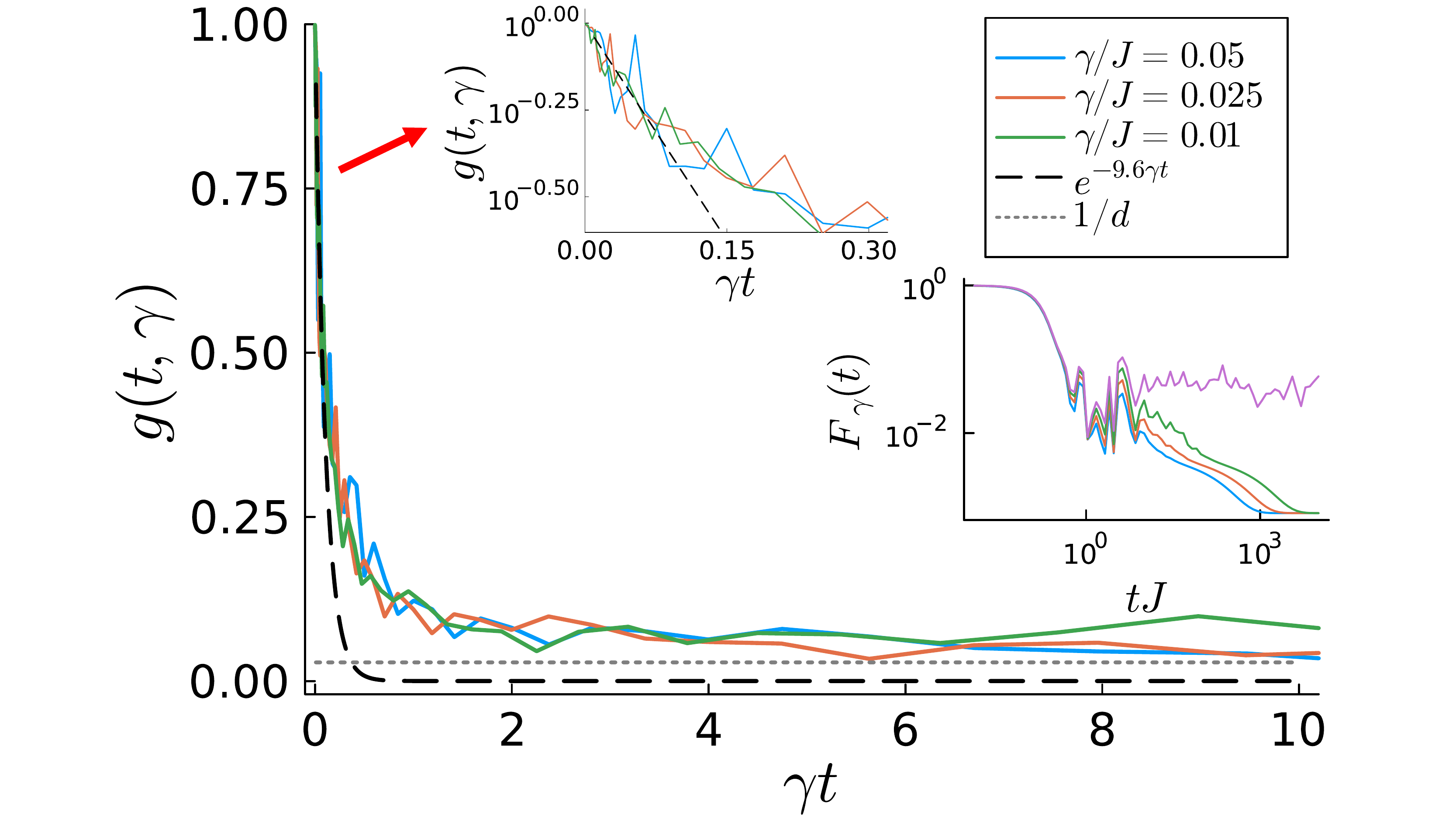}
	\caption{The normalized SFF dynamics for 1D the Bose-Hubbard model as a function of $\gamma t$. $\gamma$ is the dissipation strength. The dashed line is a fitting of the initial slope based the Eq.~\eqref{alpha_def}. The gray dot line is the prediction of the final plateau value. The left inset shows the short-time behavior of the normalized SFF. The right inset is a log-log plot of the SFF as a function of $tJ$, and the purple line is SFF without dissipation for comparison. Here, $U/J = 3.128$ and the number of sites $N_s = 4$, and the number of bosons $N_b = 4$.}
	\label{BHM_compare}
\end{figure}

The normalized SFF of the Bose-Hubbard model is illustrated in Fig.~\ref{BHM_compare}, and SFF is shown in the right inset. In our numerical simulation, we set $U/J=3.128$ which is in the quantum critical region of 1D BHM\cite{IppeiBHMOneDim2011}. Previous simulation suggest that the system exhibits quantum many-body chaos\cite{shenOutoftimeorderCorrelationQuantum2017}, although it is debatable that the system is most chaotic near the criticality\cite{PhysRevA.102.032208,pausch2022optimal}. Meanwhile, since the SFF has extensive spikes in the Bose-Hubbard model, we perform the time slice average to get a smooth SFF curve in Fig.~\ref{BHM_compare}. This time average excludes possible non-universal features, as discussed in \cite{PhysRevLett.77.1472,PhysRevLett.75.4389} for closed systems. The number of time points that we average over is $N_{average}=10$. The details of this average are added in the supplementary\cite{SM}.
The initial exponential decay curve obtained by Eq.~\eqref{alpha_def} is also included for comparison. The early-time exponential decay of the normalized SFF is visible in the left inset of Fig.~\ref{BHM_compare}, and it agrees well with the theoretical curve at $\gamma t<0.15$.

\emph{Conclusion.} In this letter, we have generalized the SFF to open quantum systems driven by the Lindblad master equation. We show that the normalized SFF of open systems generally has a dip-ramp structure and then decays to the plateau behavior at small dissipation strength. In particular, we unveil two universal properties of the normalized SFF including the early-time exponential decay behavior determined by the Lindblad operators and the late-time plateau behavior that relates to the number of steady states. Our main tools are the SYK model, the random matrix model, and the Bose-Hubbard model. Using numerical techniques, we have obtained the behavior of SFF in these three models at all times. Then we are able to extract the universal early time and late time behavior of the normalized SFF, and we find good agreement between the numerics and analytical results.

Our work potentially opens up many interesting directions: firstly, the dynamics of the SFF of open systems have a close relationship with the Lindblad spectrum\cite{ZyczkowskiUniversalSpectraRandom2019}, and therefore the SFF can be used as a diagnosis of the structure of the Lindblad spectrum. Secondly, it will be interesting to study the intermediate time scales behavior of the SFF of the open system which might go through a phase transition and have some critical behaviors\cite{kawabataDynamicalQuantumPhase2022}. Thirdly, the SFF in open systems that we discussed here can be similarly measured in experiments \cite{vasilyevMonitoringQuantumSimulators2020b,ZollerProbingManyBodyQuantum2022} via generalization to the double space, and the detail is left to the appendix\cite{SM}. 

Meanwhile, the dynamical manifestations of level repulsion can be shown in the form of a drop in the value of the survival probability below its saturation point, which is known as the correlation hole \cite{leviandierFourierTransformTool1986a,piqueChaosDynamics3001987a,guhrCorrelationsAnticrossingSpectra1990,lombardiUniversalNonuniversalStatistical1993a,torres-herreraDynamicalManifestationsQuantum2017b}. Since the survival probability is the probability of finding the system in its initial state at a later time, this survival probability is the same as the SFF at inverse temperature $\beta$ when the initial state is chosen as the coherent Gibbs state of inverse temperature $\beta$. Therefore, the dip-ramp behavior of the correlation hole has a close relation to the SFF. To generalize the study of survival probability in open quantum systems, we can replace the unitary evolution with an evolution governed by the Lindblad master equation. This results in an alternative definition of the SFF in open systems, which includes off-diagonal terms between eigenstates of the Hamiltonian, as discussed in Appendix \ref{AppendixA}. Numerically, we verify that this new definition closely corresponds to the definition discussed in the main text for nearly the entire time regime. Consequently, we anticipate that the investigation of correlation holes can also be extended to open systems. A comprehensive study of this extension, however, will be postponed in future works.

\textit{Acknowledgements.} We thank Hui Zhai for the invaluable discussions and for carefully reading the manuscript. We thank Yingfei Gu, Haifeng Tang, Liang Mao, and Hanteng Wang for the helpful discussions. We especially thank Adolfo del Campo for assisting us in rectifying our estimation of the decay exponent in the early-time regime and for bringing to our attention several relevant papers that were overlooked in a previous version.

\bibliography{ref.bib}

\appendix

\section{An alternative approach to getting the definition of the spectral form factor in open system}\label{AppendixA}
In this section, we provide another definition of the spectral form factor  (SFF) in open quantum systems whose dynamics are driven by the Lindblad master equation. Also, we compare this new definition of SFF with that we have used in the main text.
In the closed system, SFF can also be defined through the fidelity between the system's density matrix and the coherent Gibbs state
\begin{equation} 
	F(t,\beta)=\langle \psi_\beta | \rho(t) 	|\psi_\beta\rangle.
	\label{close_SFF_definition_appendix}
\end{equation}
Here, the coherent Gibbs state of inverse temperature $\beta$ is defined as
\begin{equation} \label{coherent}
	|\psi_\beta\rangle=\frac{1}{\sqrt{\mathcal{Z}(\beta)}}\sum_n e^{-\frac{\beta E_n}{2}} |n\rangle
\end{equation}
with $\mathcal{Z}(\beta)=\sum_n e^{-\beta E_n}$.
In open systems, we consider the time evolution of the system driven by the Lindblad Master equation 
\begin{equation} 
	\frac{\partial{\rho}}{\partial t}=-i[H,\rho] +\sum_{\alpha}\gamma_{\alpha}L_{\alpha} \rho L^{\dagger}_{\alpha}-\frac{1}{2} \sum_{\alpha}\gamma_{\alpha} \lbrace L^{\dagger}_{\alpha} L_{\alpha}, \rho\rbrace,
\end{equation}
and we assume that the initial state is the coherent Gibbs state, then the initial density matrix is $	\rho_{\beta}=	|\psi_\beta\rangle \langle \psi_\beta |$.
If we use the Choi-Jamiolkwski isomorphism to map the density matrix to a wave function defined on a double space
\begin{equation} 
	| \psi_{\rho}^D(t)\rangle=\sum_{mn}\rho_{mn}|m\rangle\otimes |n\rangle,
\end{equation}
then, after this mapping the wave function $\psi_{\rho}^D$ in the double system satisfies a Schrodinger-like equation
\begin{equation} 
	i\hbar \partial_t{\psi_{\rho}^D(t)}=H^D\psi_{\rho}^D(t).
\end{equation}
Here, $H^D=H_s-iH_d$ with $H_s = H_L\otimes \mathcal{I}_R - \mathcal{I}_L\otimes H^T_R$, and $H_d =\gamma (-2\hat{L}_L\otimes \hat{L}_R^*+(\hat{L}^\dagger\hat{L})_L\otimes \mathcal{I}_R +\mathcal{I}_L\otimes (\hat{L}^\dagger\hat{L})_R^*$, and operators with subscript $L$ and $R$ stand for operators acting on the left and the right systems respectively, and $T$
stands for the transpose, and $\mathcal{I}$ represents the identity operator.
Using this mapping, we can rewrite the SFF defined in Eq.~\eqref{close_SFF_definition_appendix} as
\begin{equation} 
	\tilde{F}_\gamma(t,\beta)=\langle \psi_\beta^D | \psi_{\rho}^D(t)\rangle.
\end{equation}
Here, we use the subscript $\gamma $ to denote the SFF in the open system (that is the dissipation strength is non-zero). Then, SFF can be viewed as the overlap between the double space wave function at time $t$ and the double space initial wave function, which is the double space coherent Gibbs state defined as $|\psi_\beta^D \rangle= 	|\psi_\beta\rangle \otimes 	|\psi_\beta\rangle$. This way of defining SFF in open system was first introduced in \cite{xuThermofieldDynamicsQuantum2021} where they have defined the SFF as the fidelity between the initial (pure) thermofield double (TFD) state and its evolution.

We consider that this non-Hermitian Hamiltonian $H^D$ can be diagonalized, and gives a set of eigenstates that satisfy $H^D|\psi_{\rho}^l(t)\rangle=\epsilon_l |\psi_{\rho}^{D,l}(t)\rangle$ with $\epsilon_l=\alpha_l+i\beta_l$ is the eigenvalue of the eigenstate $l$.
The initial state in the double space can be expanded as 
\begin{equation} 
	| \psi_{\rho}^D(0)\rangle=	|\psi_\beta^D \rangle= \sum_l c_{l} | \psi_{\rho}^{D,l}\rangle,
\end{equation}
and the time evolution of this double space wave function is given by
\begin{equation} 
	| \psi_{\rho}^D(t)\rangle=\sum_l c_{l} e^{(-i\alpha_l-\beta_l)t}| \psi_{\rho}^{D,l}\rangle.
\end{equation}
Here, $c_{l} =\langle \psi_{\rho}^{D,l}|\psi_\beta^D \rangle$. Therefore, the SFF can be further written as
\begin{equation} 
	\tilde{F}_{\gamma}(t,\beta)=\sum_l c_{l} e^{(-i\alpha_l-\beta_l)t}\langle \psi_\beta^D | \psi_{\rho}^{D,l}\rangle=\sum_l |c_{l}|^2 e^{(-i\alpha_l-\beta_l)t}.
	\label{sff_closed}
\end{equation}
Thus, we find Lindblad spectrum and the initial state distribution on the Lindblad spectrum fully determine this quantity. Since $\beta_l$ is always non-negative, SFF will not grow exponentially. Consequently, although the Lindblad spectrum is complex, the SFF calculated from it will decay exponentially in time till it reaches its steady-state value. For the steady-state, the plateau value of the SFF is given by
\begin{equation} 
	\lim_{t\to \infty}	\tilde{F}_{\gamma}(t,\beta)=\sum_{\beta_0=0,\alpha_0}|c_{0}|^2 e^{-i\alpha_0t}.
\end{equation}
Thus, the plateau value of the SFF depends on the overlap of the initial state and the steady state.

\begin{figure}[t] 
	\centering 
	\includegraphics[width=0.45\textwidth]{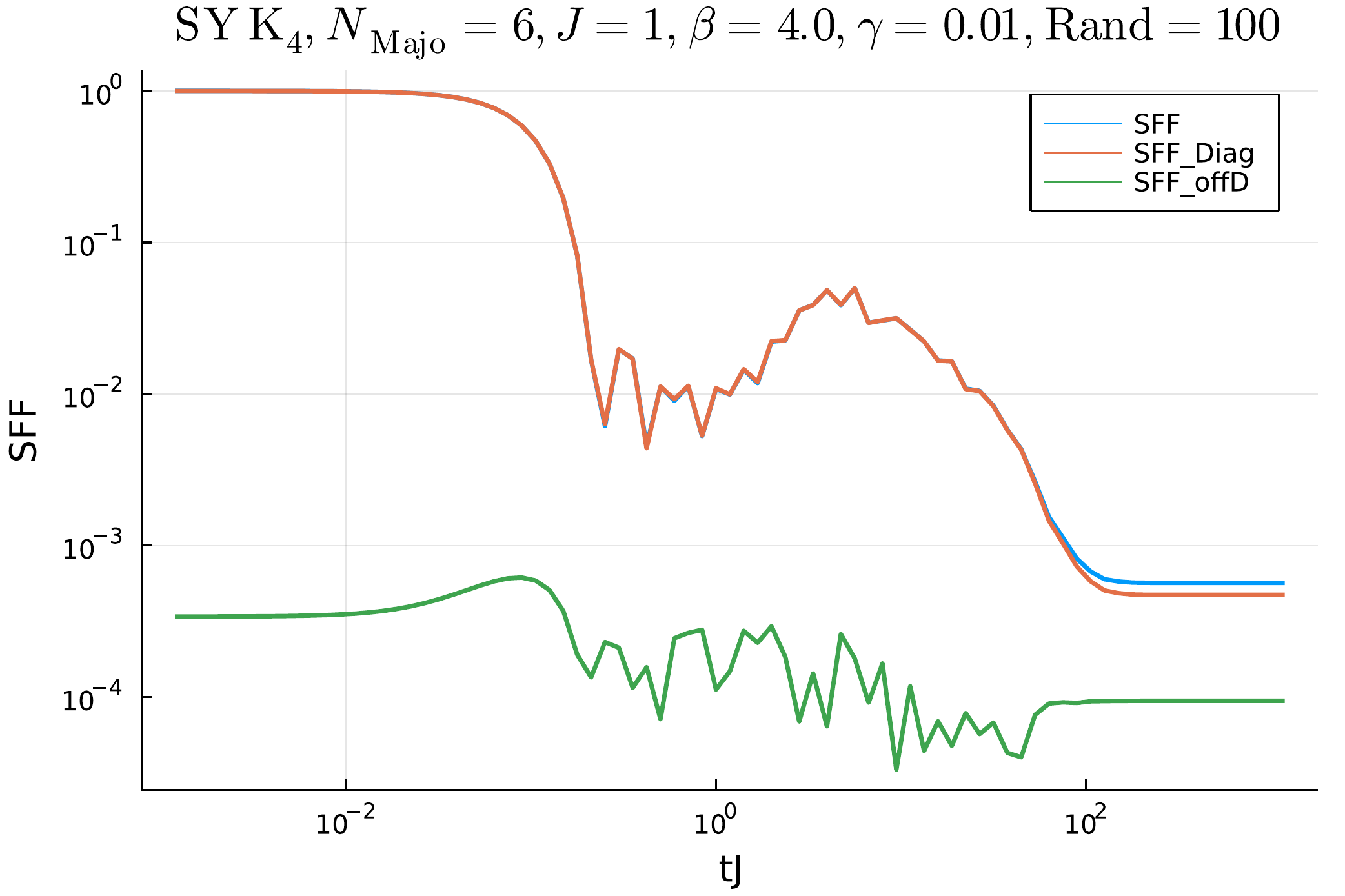} 
	\caption{The comparison of two different definitions of SFF in the open system. The red line is the SFF defined in Eq.~\eqref{definition_1_appendix}, and the blue line is the SFF defined in Eq.~\eqref{definition_2_appendix}. The green line is the difference between these two definitions.}
	\label{2def}
\end{figure}

In addition, since the second Renyi entropy is defined as $	e^{-s^{(2)}}=\langle \psi_{\rho}^D(t)| \psi_{\rho}^D(t)\rangle$, there is a direct connection between the dynamics of the SFF and entropy dynamics. The SFF defined above can be expressed in the energy basis as
\begin{equation} 
	\begin{split}
		\tilde{F}_{\gamma}(t,\beta)  =&\frac{1}{[\mathcal{Z}(\beta)]^2}\sum_{m,n,m',n'}e^{-\frac{\beta}{2} (E_m+E_n+E_{m'}+E_{n'})} \\
	&\langle m|\otimes \langle n|e^{-iH^Dt} |m'\rangle \otimes |n'\rangle.
	\end{split}
	\label{definition_1_appendix} 
\end{equation}
Here, $m,n,m',n'$ are the eigenstates of $H$. We can further define a new type of SFF by preserving only the diagonal elements in this eigenstate basis of the Eq.~\eqref{definition_1_appendix}. This new definition of SFF can be written explicitly as 
\begin{equation} 
	F_{\gamma}(t,\beta)=\frac{1}{[\mathcal{Z}(\beta)]^2}\sum_{m,n} e^{-\beta (E_n+E_m)} \langle m|\otimes \langle n|e^{-iH^Dt} |m\rangle \otimes |n\rangle.
	\label{definition_2_appendix}
\end{equation}
And this is the definition of the SFF in the open system that we have used in the main text if we set $\beta=0$. We can then numerically compare these two different definitions of SFF in the open system,  as shown in  Fig.~\ref{2def}. We find that these two definitions are almost the same in the SYK model as an example.

If we consider the case of the infinite temperature $\beta=0$, we have 
\begin{equation} 
	\begin{split}
		F_\gamma(t,\beta=0) =&\frac{1}{[\mathcal{Z}(0)]^2} \Tr(e^{-iH^D t})\\
		=&\frac{1}{[\mathcal{Z}(0)]^2}\sum_l   e^{(-i\alpha_l-\beta_l)t}  
	\end{split}	
	\label{definition_2_beta0_appendix}
\end{equation}
Let us now compare these definitions of SFF of an open system in Eq.~\eqref{definition_2_beta0_appendix} and that in Eq.~\eqref{sff_closed}. We find that this new definition in the Eq.~\eqref{definition_2_beta0_appendix} simply takes all $|c_{l}|^2=\frac{1}{[\mathcal{Z}(0)]^2}$ in the Eq.~\eqref{sff_closed}.

\section{The derivation of the pre-factor $\alpha$ in early decay}
In this section, we give the detailed derivation of the expression of pre-factor $\alpha$ in the early decay region.
At small $\gamma t$, the SFF becomes
\begin{equation} 
	\begin{split}
		F_\gamma(t)&= \frac{1}{[\mathcal{Z}(0)]^2}\operatorname{Tr}[  e^{-iH^Dt}]\\
		&\simeq \frac{1}{[\mathcal{Z}(0)]^4}\operatorname{Tr}[e^{-iH_st}]\operatorname{Tr}[e^{-H_dt}].
	\end{split}
\end{equation}
Using the definition of $H_s$
\begin{equation} 
	H_s = H_L\otimes \mathcal{I}_R - \mathcal{I}_L\otimes H^T_R,
\end{equation}
we obtain 
\begin{equation} 
	\begin{split}
		F_\gamma(t)\simeq \frac{1}{[\mathcal{Z}(0)]^2} F(t,\beta=0)\operatorname{Tr}[e^{-H_dt}].
	\end{split}
	\label{initial_alpha_appendix}
\end{equation}
Now, recall the definition of $H_d$
\begin{equation} 
\begin{aligned}
	H_d =&\gamma \sum_m [-2\hat{L}_{m,L}\otimes \hat{L}_{m,R}^*+(\hat{L}^\dagger_m\hat{L}_m)_L\otimes \mathcal{I}_R \\&+\mathcal{I}_L\otimes (\hat{L}^\dagger_m\hat{L}_m)_{R}^*].
	\label{Hd_appendix}
 \end{aligned}
\end{equation}
Keeping to the leading order in $\gamma t$, we find
\begin{equation} 
\begin{aligned}
	\frac{\operatorname{Tr}[e^{-H_dt}]}{{[\mathcal{Z}(0)]^2} }\approx &\frac{1}{{[\mathcal{Z}(0)]^2} }\operatorname{Tr}[1-H_dt]\\
 &=1-2 \gamma t\sum_m[2\langle\hat{L}^\dagger_m\hat{L}_m\rangle-|\langle \hat{L}_m\rangle|^2].\\
 &\approx e^{-2 \gamma t\sum_m[2\langle\hat{L}^\dagger_m\hat{L}_m\rangle-|\langle \hat{L}_m\rangle|^2]}
 \end{aligned}
\end{equation}
Here we have used the fact that $\mathcal{Z}(0)=d$ is the Hilbert space dimension of the Hamiltonian $H$. Finally, we obtain
\begin{equation} 
	g(t,\gamma)  \equiv\frac{	F_\gamma(t) }{F(t,\beta=0) }=e^{-\alpha \gamma t}.
\end{equation}
with 
\begin{equation} 
	\alpha = \sum_m 2\langle\hat{L}^\dagger_m\hat{L}_m\rangle_c.
\end{equation}
Here $\langle A\rangle\equiv \operatorname{Tr}\left[ A\right]/d$ and $\langle A B\rangle_c \equiv \langle A B\rangle-\langle A \rangle\langle B\rangle$.

\section{SFF of the random matrix theory}

\subsection{Random matrix theory in closed system}
In this section, we review some basics of calculating SFF in the random matrix theory (RMT). Let us first review how to calculate the SFF in closed systems in RMT. The $N \times N $ random Hermitian matrix $H$ of the Gaussian Unitary Ensemble (GUE) is defined to be averaged for the following Gaussian distribution:
\begin{equation} 
	P(H)=\frac{1}{\mathcal{Z}}\exp[-\frac{N}{2}\Tr(H^2)].
\end{equation}
One can write this Gaussian distribution in the eigenvalue basis, where the distribution over the set of matrices could reduce to the distribution of eigenvalues with the following joint distribution
\begin{equation} 
	P(\lambda_1,\lambda_2,...,\lambda_N)=\exp[-\frac{N}{2}\sum_{i=1}^N \lambda_i^2]\prod_{i<j}^{N}(\lambda_i - \lambda_j)^2.
\end{equation}
We could further compute the $n$-point correlation function ($n<N$) as
\begin{equation} 
	\rho^{(n)}(\lambda_1,\lambda_2,...,\lambda_n)= \int d\lambda_{n+1}...\lambda_{N} P(\lambda_1,\lambda_2,...,\lambda_N).
\end{equation}
People find that the correlation function could be determined by a kernel $K$ in the large N limit\cite{wignerDistributionRootsCertain1958,RandomMatricesVolume,guhrRandommatrixTheoriesQuantum1998a}:
\begin{equation} 
	\rho^{(n)}(\lambda_1,\lambda_2,...,\lambda_n)= \frac{(N-n)!}{N!}\det(K(\lambda_i,\lambda_j))_{i,j=1}^n
\end{equation}
where the kernel $K$, in the large $N$ limit, behaves as
\begin{equation} 
	K(\lambda_i,\lambda_j)= \begin{cases}
		\frac{N}{2\pi} \sqrt{4-\lambda_i^2}, & i=j  \\
		\frac{N}{\pi} \frac{\sin[L(\lambda_i-\lambda_j)]}{L(\lambda_i-\lambda_j)}, & i \neq j.\\
	\end{cases}
	\label{kernal_RMT}
\end{equation} 
In the colliding case $i = j$, this kernel is the familiar Wigner's semicircle law. While in the case where $i \neq j$, this kernel is called the sine kernel in RMT. Then, we can calculate the simple one-point form factor as 
\begin{equation} 
	\begin{split}
		g^{(1)}(\beta,t)
		=&\frac{1}{N}\int d\lambda_{1}\lambda_2...\lambda_{N} P(\lambda_1,\lambda_2,...,\lambda_N)\exp[-(\beta+it)\lambda_{1}]\\
		=&\frac{1}{N}\int d\lambda_{1} 	\rho^{(1)}(\lambda_1)\exp[-(\beta+it)\lambda_{1}].
	\end{split}
\end{equation}
Similarly, we can calculate the two-point form factor, which is what we called SFF in closed systems: 
\begin{equation} 
	\begin{split}
		F(t,\beta)=&g^{(2)}(\beta,t)\\
		=&\frac{1}{N^2}\int d\lambda_{1}\lambda_2...\lambda_{N} P(\lambda_1,\lambda_2,...,\lambda_N)\\
		&\exp[-(\beta+it)\lambda_{1}]\exp[-(\beta-it)\lambda_{2}]\\
		=&\frac{1}{N^2}\int d\lambda_{1} d\lambda_{2} 	\rho^{(2)}(\lambda_1,\lambda_2)\\
		&\exp[-(\beta+it)\lambda_{1}]\exp[-(\beta-it)\lambda_{2}].
		\label{SFF_RMT_closed}
	\end{split}
\end{equation}
For simplicity, we consider infinite temperature $\beta=0$,
\begin{equation} 
	\label{SFF_closed}
	F(t,\beta=0)=\frac{1}{N^2}\int d\lambda_{1} d\lambda_{2} 	\rho^{(2)}(\lambda_1,\lambda_2)\exp[-it(\lambda_{1}-\lambda_{2})].
\end{equation}
We find here that the SFF in the closed system is determined by the two-point correlation function. In the case of the infinite temperature, SFF is simply the Fourier transform of the two-point correlation function.

\subsection{Random matrix theory in open system} 
In this section, we calculate the normalized SFF of the open system in the GUE. The SFF we defined in Eq.~\eqref{definition_2_beta0_appendix} can be written in the RMT as 
\begin{equation} 
	F_\gamma(t)=\frac{1}{N^2}\langle \operatorname{Tr} [e^{iH^Dt}] \rangle
\end{equation}
Here, $H^D=H_s -i H_d$ is the random matrix defined in the double space. Here, $H_s = H_L\otimes \mathcal{I}_R - \mathcal{I}_L\otimes H^T_R$, and $H_d =\gamma (-2\hat{L}_L\otimes \hat{L}_R^*+(\hat{L}^\dagger\hat{L})_L\otimes \mathcal{I}_R +\mathcal{I}_L\otimes (\hat{L}^\dagger\hat{L})_R^*$ with $H$ and $L$ both are $N \times N $ random Hermitian matrix. The bracket $\langle * \rangle$ means an averaging for the Gaussian distribution:
\begin{equation} 
	P(H)=\frac{1}{\mathcal{Z}}\exp[-\frac{N}{2}\Tr(H^2)]
\end{equation}
and 
\begin{equation} 
	P(L)=\frac{1}{\mathcal{Z}}\exp[-\frac{N}{2}\Tr(L^2)].
\end{equation}
To understand the behavior of SFF with dissipation, we first use an approximation
\begin{equation} 
	\Tr [e^{-itH^D}] \simeq \Tr[ e^{-itH_s}e^{-tH_d}] \simeq \frac{1}{N^2}\Tr [e^{-itH_s}] \Tr[ e^{-tH_d} ].
\end{equation}
This approximation is good in the early-time limit. Then, the SFF can be formulated as
\begin{equation} 
	\begin{split}
		F_\gamma(t)\simeq&\frac{\int dH dLe^{-\frac{N}{2}\Tr H^2}e^{-\frac{N}{2}\Tr L^2}\frac{1}{N^2}\Tr e^{-itH_s}\Tr e^{-tH_d}}{\int dH dL e^{-\frac{N^2}{2}\Tr H^2}e^{-\frac{N}{2}\Tr L^2}}\\
		=&\frac{1}{ N^4}\sum_{i,j,s,r=1}^N\int d\lambda_{i}\lambda_j dl_s dl_r \rho_H^{(2)}(\lambda_i,\lambda_j)\rho_L^{(2)}(l_s,l_r)\\
		&e^{-it(\lambda_{i}-\lambda_{j})}e^{-\gamma t(l_{s}-l_{r})^2}
	\end{split}
\end{equation}
Then, we further obtain
\begin{equation}
	\begin{split}
		g(t,\gamma)=&\frac{1}{N^2}\sum_{s,r=1}^N\int dl_s dl_r\rho_L^{(2)}(l_s,l_r)e^{-\gamma t(l_{s}-l_{r})^2}\\
		=&\frac{1}{N}+\frac{(N-1)}{N}\int dl_1 dl_2\rho_L^{(2)}(l_1,l_2)e^{-\gamma t(l_{1}-l_{2})^2}.
	\end{split}
\end{equation}
Then, we calculate the normalized SFF below. Using 
\begin{equation} 
	\rho^{(2)}(l_1,l_2)=\frac{N}{(N-1)}\rho(l_1) \rho(l_2)-\frac{N}{(N-1)}\frac{\sin^2[N(l_1-l_2)]}{[N(l_1-l_2)]^2},
\end{equation}
we obtain
\begin{equation} 
	g(t,\gamma)=g(t,\gamma)^{\text{disc}}+g(t,\gamma)^{\text{conn}}.
\end{equation}
Here, the connected part of $g(t,\gamma)$ is
\begin{equation} 
	g(t,\gamma)^{\text{conn}}=\frac{1}{N}- \int dl_1 dl_2\frac{\sin^2[N(l_1-l_2)]}{[N(l_1-l_2)]^2}e^{-\gamma t(l_{1}-l_{2})^2}.
\end{equation}
We further define $u_1= l_1-l_2$ and $u_2=l_{2}$, and we use the box approximation to deal with this divergent integral. Then, we have
\begin{equation} 
	\begin{split}
		g(t,\gamma)^{\text{conn}}=&\frac{1}{N}- \int du_1 du_2\frac{\sin^2(Nu_1)}{(Nu_1)^2}e^{-\gamma tu_1^2}\\
		=&\frac{1}{N}-2u_{\text{cut}} \int du_1 \frac{\sin^2(Nu_1)}{(Nu_1)^2}e^{-\gamma tu_1^2}\\
		=&\frac{1}{N}-\frac{2u_{\text{cut}}}{N} \int dx \frac{\sin^2(x)}{(\pi x)^2}e^{-\gamma \frac{t}{L^2}x^2}
		\label{g_conn}
	\end{split}
\end{equation}
$u_{\text{cut}}$ is the constant introduced with box approximation. Besides, the disconnected part of $g(t,\gamma)$ is
\begin{equation} 
	\begin{split}
		g(t,\gamma)^{\text{disc}}=&\int dl_1 dl_2\rho(l_1) \rho(l_2)e^{-\gamma t(l_{1}-l_{2})^2}\\
		=&\int_{-2}^2 dl_1\int_{-2}^2  dl_2\frac{1}{(2\pi)^2}\sqrt{4-l_1^2}\sqrt{4-l_2^2}e^{-\gamma t(l_{1}-l_{2})^2}.
	\end{split}
\end{equation}

\begin{figure}[t] 
	\centering 
	\includegraphics[width=0.45 \textwidth]{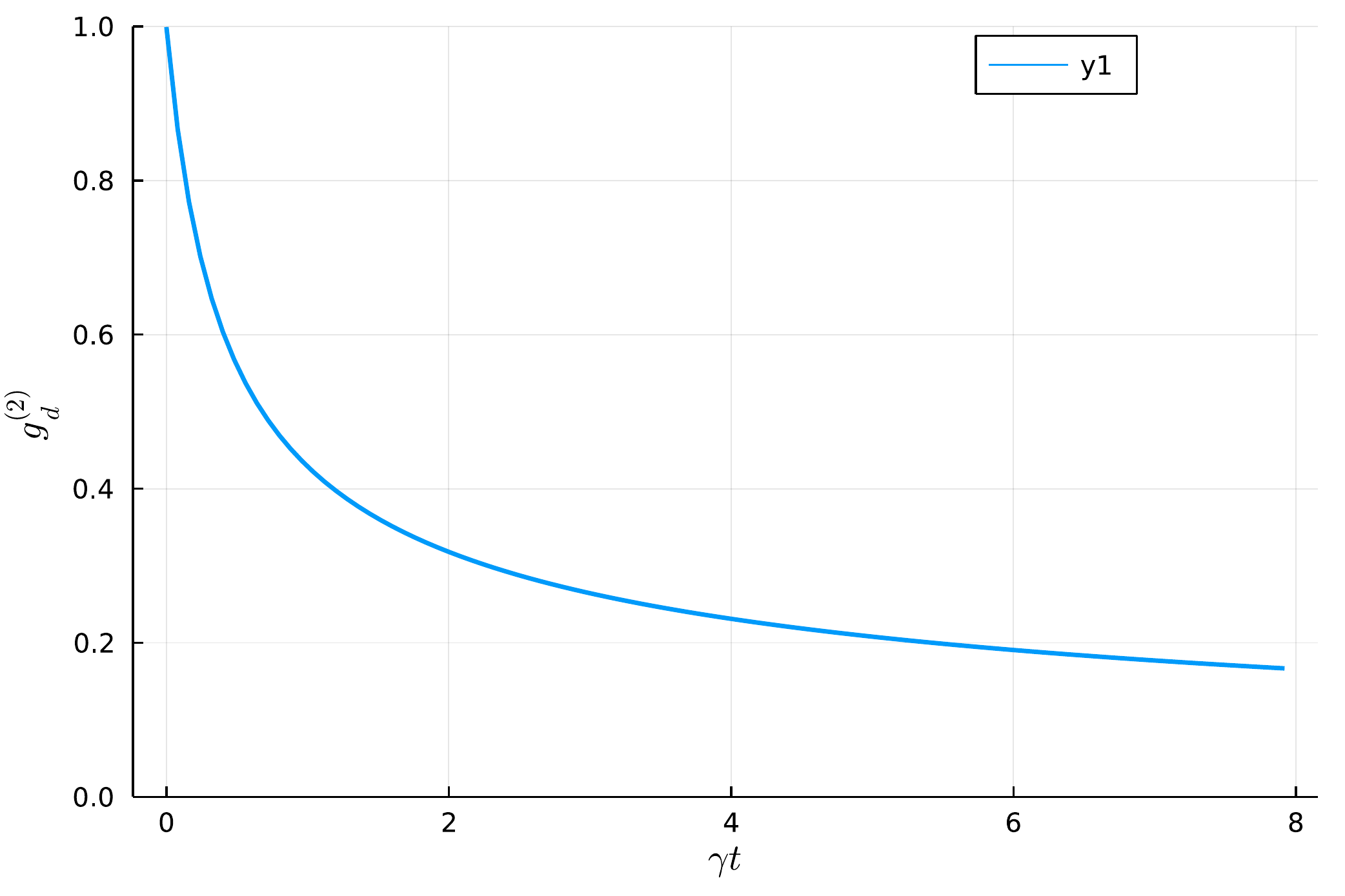} 
	\caption{The numerical results for $g(t,\gamma)$ dynamics in the Eq.~\eqref{g2_integral} at the limit $N \to \infty$.} 
	\label{g_disconnect}
\end{figure}

One can try to solve the $u_{\text{cut}}$ in the Eq.~\eqref{g_conn} by checking the consistency of the result at $t = 0$. We notice that normalization condition gives 	$g(0,\gamma)=1$, and direct calculation shows $g(0,\gamma)^{\text{disc}}=1$. These two facts lead to $g(t,\gamma)^{\text{conn}}=0$. As a result, the connected part can be determined as
\begin{equation} 
	g(t,\gamma)^{\text{conn}}=\frac{1}{N}- \frac{1}{N\sqrt{\gamma \frac{t}{N^2}+\sqrt{1+(\gamma \frac{t}{N^2})}^2}}\simeq \frac{\gamma t}{2N^3}.
\end{equation}

\section{SFF of SYK model}
\subsection{The Brownian SYK with dissipation}
We consider the SFF in the Brownian SYK model with dissipation. Its Hamiltonian is of the form
\begin{equation} 
	H(t) = i^{\frac{q}{2}}\sum_{a_1<...<a_q}^N J_{a_1,...,a_q}(t)\psi_{a_1}...\psi_{a_q}.
\end{equation}
Here, $J_{a_1,...,a_q}(t)$ is a random variable that satisfies the Gaussian
distribution with zero mean value and variance  
\begin{equation} 
	\langle J_{a_1,...,a_q}(t) J_{a_1^{'},...,a_q^{'}}(t') \rangle=\delta_{a_1,a_1^{'}}...\delta_{a_q,a_q^{'}}\delta(t-t')\frac{J(q-1)!}{N^{q-1}}.
\end{equation}
We first write the SFF of the Brownian SYK model as a path integral with the Lindblad operator chosen as the single Majorana operator $L_i = \psi_i$. Also, the dissipation strength is chosen as the constant $\gamma$. Then, the SFF in the open system can be written as 
\begin{equation}
	\begin{split}
	F_\gamma(T) &=\frac{1}{2^N}\langle \Tr e^{-itH^D}\rangle\\
	&=\frac{1}{2^N}\int \mathcal{D}\psi^L_a \mathcal{D}\psi^R_a  \exp\Bigg\{ i \int_0^T dt\Big[ \frac{i}{2}\psi_a^{(j)}\partial_t\psi_a^{(j)}\\&-J_{a_1 a_2 ... a_q}(t) 
	(i^{\frac{q}{2}}\psi^{L}_{a_1 a_2 ... a_q}-(-i)^{\frac{q}{2}}\psi^{R}_{a_1 a_2 ... a_q} )  \\& +2\gamma \psi_a^L  \psi_a^R-iN\gamma     \Big] \Bigg\}.
\end{split}
\end{equation}
After integrating out $J_{a_1 a_2 ... a_q}(t)$ variables, we obtain
\begin{equation}
	\begin{split}
		F_\gamma(T) =&\frac{1}{2^N}\int \mathcal{D}\psi^L_a \mathcal{D}\psi^R_a \exp\Bigg\{ -\int_0^T dt \Big[ \frac{1}{2}\psi_a^{(j)}\partial_t\psi_a^{(j)} \\&+\frac{J^2(q-1)!}{N^{q-1}}\sum_{a_1<a_2<...<a_q}(\frac{1}{2^q}-\psi^{L}_{a_1 a_2 ... a_q} \psi^{R}_{a_1 a_2 ... a_q} )\\&+2i\gamma \psi_a^L  \psi_a^R+N\gamma  \Big] \Bigg\}.
\end{split}
\end{equation}
Furthermore, we can represent Majorana fermions in terms of spin variables as $\psi_a^L  \psi_a^R=\frac{i}{2}\sigma_a^z$, then the above expression can be understood as a normal thermal partition function of a spin system $F_\gamma(t,\beta=0) =\frac{1}{2^N}\exp[-TH_{spin}]$ with 
\begin{equation} 
	H_{spin}=\frac{J^2(q-1)!}{2^qN^{q-1}}\sum_{a_1<a_2<...<a_q}(1-\sigma_1^z...\sigma_q^z)-\gamma\sum_a\sigma_a^z+N\gamma.
\end{equation}

In the long time limit, this factor can be viewed as a projection operator onto the ground states. The two ground states are when all spins are up or down without dissipation. When we add small dissipation as a perturbation to the original degenerate ground stats, the energy of the two perturbed ground states is $ E_g = 0$ and $2N\gamma$. Following the argument, the SFF reads 
\begin{equation} 
	\begin{split}
		F_\gamma(t\to \infty,\beta=0) =&\frac{1}{2^N}\left\{\exp[-T\times0]+\exp[-T\times2N\gamma]\right\}\\
		=&\frac{1}{2^N}\left\{1+\exp[-2N\gamma T]\right\}.
	\end{split}
\end{equation}
\begin{figure*}
	\includegraphics[width=\textwidth]{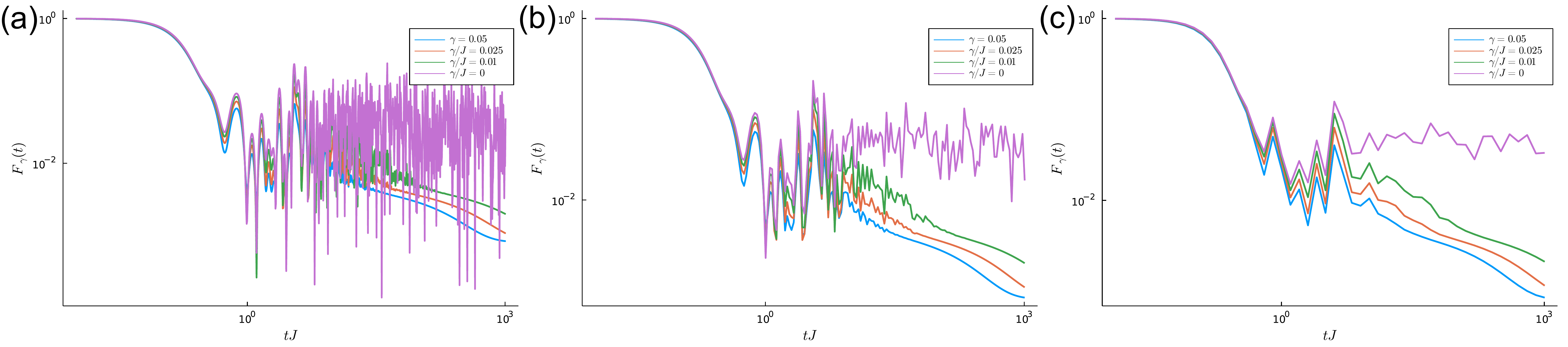}
	\caption{The log-log plot of the SFF as a function of $tJ$, and the purple line is SFF without dissipation for comparison. Here, the time slice average $N_{average}=1,5,20$ for (a),(b),(c) respectively.}
\label{BHM}
\end{figure*}

At long time limit $\gamma T \gg 1$, this SFF is $1/2^N$ which is half of that with no dissipation. This explains the late-time plateau behavior of the SFF in the SYK model with dissipation, and the plateau value is $1/2$ of that with zero dissipation. This is because the dissipation breaks the degeneracy of the ground states, and gives a positive energy correction to one of the original ground states, thus this state decays as time increases, and only one state with zero energy survives.  

Next, we look for the saddle points in the semi-classical analysis of this SFF. There is an exact familiar rewrite of the SYK model in terms of variables $G$ and $\Sigma$. The path integral of SFF can be rewritten in terms of  variables $G_{LR}$ and $\Sigma_{LR}$
	\begin{equation}
 \begin{aligned}
		F_\gamma(T) =&\frac{1}{2^N}\int \mathcal{D}G_{LR} \mathcal{D}\Sigma_{LR} \mathcal{D}\psi^L_a \mathcal{D}\psi^R_a \exp\left\{ -S_F\right\}.\\
  S_F=&\int_0^T dt~\frac{N}{2} \Bigg[ \frac{2J}{q}(\frac{1}{2^q}-i^q G_{LR}^q )-4\gamma G_{LR}+2\gamma\\&+\Sigma_{LR} G_{LR}  \Bigg] + \frac{1}{2}\left[\psi_a^{(j)}\partial_t\psi_a^{(j)} - \psi_a^{L}\psi_a^{R}\Sigma_{LR}\right].
\end{aligned}
\end{equation}

 Without dissipation, we know an obvious saddle point which is simply $G_{LR}=\Sigma_{LR}=0$. Then, we assume that there is a saddle point near 0 when $\gamma t \ll 1$.
The total $G_{LR},\Sigma_{LR}$  integrand becomes
\begin{equation}
\begin{aligned}
	\exp\Bigg\{ N\Bigg[\log(2\cos(\frac{T\Sigma_{LR}}{4}))-\frac{JT}{q2^q}+i^q\frac{JT}{q}G_{LR}^q\\
 -\frac{T}{2}\Sigma_{LR}G_{LR}-2i\gamma TG_{LR}-\gamma T\Bigg]\Bigg\}.
 \end{aligned}
\end{equation}
We assume $G_{LR},\Sigma_{LR}$ to be very small, thus $\tan(\frac{T\Sigma_{LR}}{4})\simeq \frac{T\Sigma_{LR}}{4}$. We here simply take $q=2$, then we arrive at the solution 
\begin{equation} 
	\Sigma_{LR}=-4i\gamma, G_{LR}=\frac{i\gamma T}{2}.
\end{equation}
Then, the saddle point action according to this saddle point is $	2^N\exp[-\frac{JNT}{8}]\exp[-N\gamma T]$, and this explains the early-time exponential decay behavior of the SFF in Brownian SYK model. In addition, there is another non-trivial saddle points solution. 
We can obtain one set of saddle points in the long-time limit
\begin{equation} 
	G_{LR}=\pm \frac{i}{2}, \Sigma_{LR}=\mp \frac{iJ}{2^{q-2}}-4i\gamma.
\end{equation}
Then, the saddle point action is $0, -2N\gamma T$. Therefore, the SFF is $	\exp[0]+\exp[-2N\gamma T]=1+\exp[-2N\gamma T]$, and this saddle point explains the late-time plateau behavior of SFF. This is exactly what we obtain through the spin Hamiltonian analysis.

\subsection{The regular SYK with dissipation}
We further consider the normal SYK model. Similar to the Brownian SYK case, we can write a path-integral expression for the SFF of the regular SYK model as
\begin{equation} 
	F_\gamma(t) =\int  \mathcal{D}G \mathcal{D}\Sigma e^{-NI[G,\Sigma]}  
\end{equation}
with	
\begin{equation}
\begin{aligned}
	I[G,\Sigma]&=-\log Pf(\delta_{ab}\partial_t-\Sigma)+\gamma T+\frac{1}{2}\int_0^T\int_0^Tdt_1 dt_2\\
 &\lbrack\Sigma_{ab}G_{ab}-\frac{J^2}{q}s_{ab}G^q_{ab}+2i\gamma G_{ab}\delta_{ab^{'}}\delta(t_1-t_2)\rbrack.
 \end{aligned}
\end{equation}
Taking the saddle-point equation, to the first order of $\gamma$, we have  
\begin{equation}
\begin{aligned}
		\binom{G_{LL}\ G_{LR}}{G_{RL}\ G_{RR}}\simeq&-\frac{1}{(i\omega+\Sigma_{LL})(i\omega+\Sigma_{RR})}\times\\&
  \dbinom{iw+\Sigma_{RR} \ \ \  -\Sigma_{LR}}{-\Sigma_{RL} \ \ \  iw+\Sigma_{LL}}
  \end{aligned}
\end{equation}
We take $q=2$ for simplicity, and we have
\begin{equation} 
	\begin{split}
		&G_{LL}=G_{RR}=\frac{i\omega\pm\sqrt{4J^2-\omega^2}}{2J^2}\\ &G_{LR}=G_{RL}=\frac{-2i\gamma G_{LL} G_{RR} }{1+J^2G_{LL} G_{RR}}.
	\end{split}
\end{equation}
The saddle point action is $	I[G,\Sigma]=I_0[G,\Sigma]+\gamma T$. Thus, we obtain the normalized SFF as $g(t,\gamma) =\exp[-N\gamma T]$. It has an exponential decay behavior at early time.

\section{SFF of Bose-Hubbard model}
The SFF of the Bose-Hubbard model has extensive spikes that come from the zeros of the SFF. In order to transform the curve of SFF to a smooth function, we perform a time average to the numerical results of the SFF for the Bose-Hubbard model, as we treat in Fig.~(4) of the main text. From time $tJ$ between $10^{-1}$ to $10^3$, we equally divide the total time into $N_t=1000$ pieces in the log scale. And we get the SFF at each time point by averaging the value of SFF between $N_{average}=10$ neighbor points. Also, the larger the $N_{average}$ becomes, the smoother the SFF curves will be. Below, we show the numerical results of SFF at different $N_{average}=1,5,20$ in Fig.~\eqref{BHM}.

\section{The experimental realization of the SFF in open systems.}
\begin{figure}[t] 
	\centering 
	\includegraphics[width=0.45\textwidth]{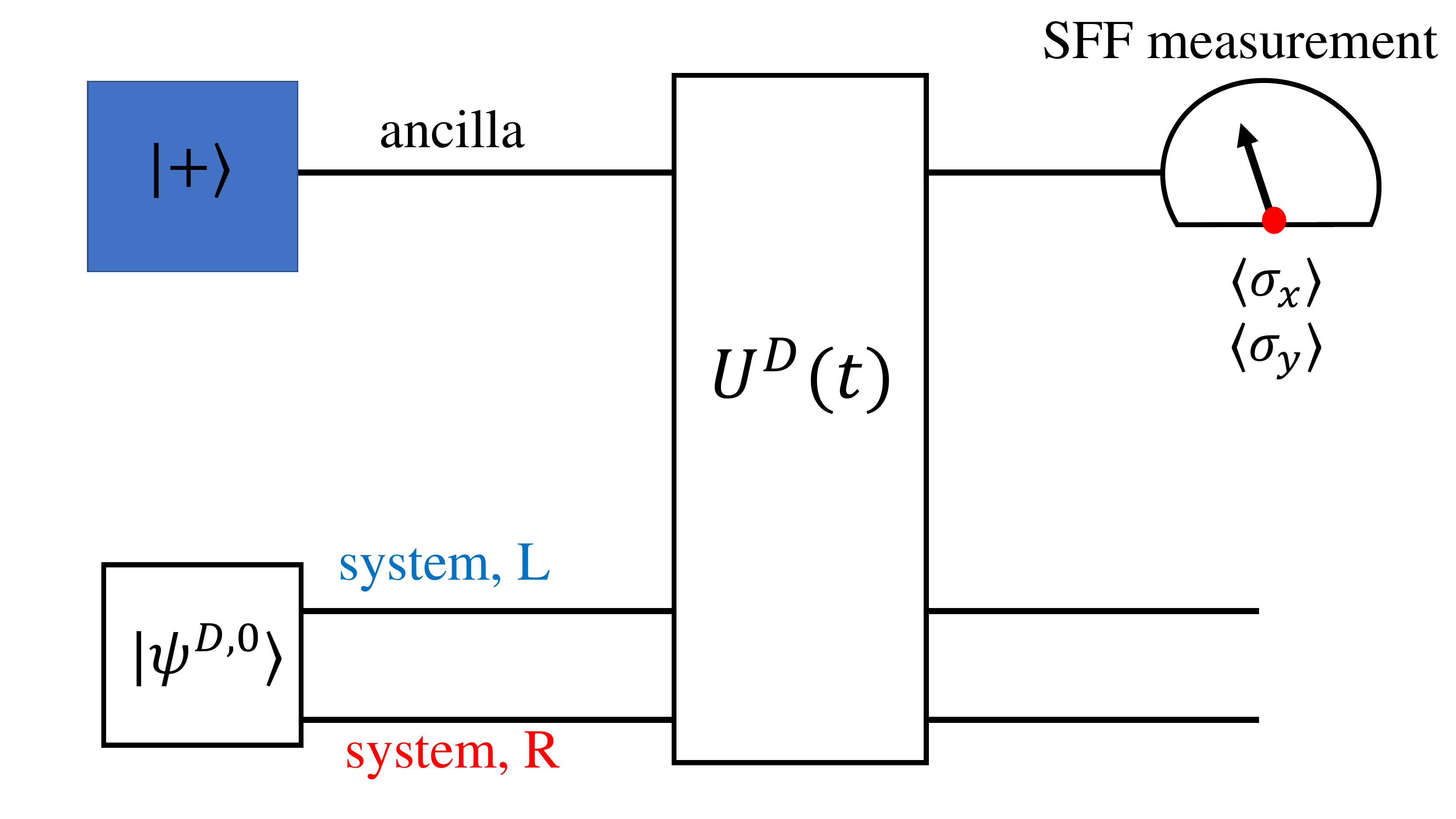}
	\caption{A quantum circuit employing a QND coupling of the quantum simulator in double space to an ancilla qubit to measure the SFF.} 
	\label{experiment}
\end{figure}

In this section, we give a possible experiment realization proposal of the SFF in open systems. We first prepare initial the double space wave function as
\begin{equation} 
	|\psi^{D,0}\rangle = \frac{1}{\sqrt{N}}\sum_n|n\rangle_L\otimes|n\rangle_R,
\end{equation}
then we perform the evolution for a time $t$ with the
quantum non-demolition (QND) Hamiltonian in the double space 
\begin{equation} 
	\mathcal{U}^D_{QND}(t)=\exp\left[-iH^D_{QND}(t)\right]
\end{equation}
with 
\begin{equation} 
	H^D_{QND}=H^D\otimes|0\rangle_{c} \langle 0|.
\end{equation}
Here, $H^D$ is the mapping of the Lindblad master equation onto the double space, and 'c' denotes the ancilla qubit which is also called the control qubit. 
Finally, we measure the expectation
values of $\sigma_x$ and $\sigma_y$ for the ancilla qubit, as shown in Fig.~\eqref{experiment}.
After direct calculation, we obtain 
\begin{equation} 
	\langle \sigma_x(t)\rangle=\sum_l \cos(\epsilon_lt),
\end{equation}
and
\begin{equation} 
	\langle \sigma_y(t)\rangle=\sum_l \sin(\epsilon_lt).
\end{equation}
Here, $\{\epsilon_l\}$ is the Lindblad spectrum. Therefore, SFF can be obtained by 
\begin{equation} 
	F_{\gamma}(t)=\langle \sigma_y(t)\rangle-i\langle \sigma_x(t)\rangle.
\end{equation}

\end{document}